\documentclass[twoside]{LCWS11}
\usepackage[latin1]{inputenc}
\usepackage[dvips]{graphicx,epsfig,color}
\usepackage{wrapfig,rotating}
\usepackage{amssymb,amsmath,array}
\usepackage{cite}
\begin{document}
\def\d#1{D_{#1}}
\def\db#1{\bar D_{#1}}
\title{
%%%%   Paper title goes here  %%%%%%%%%%%%%%
Status of {\sc MadLoop}/a{\sc MC@NLO}
} %% 
%***********************************************************************
% AUTHORS INFORMATION AREA
%***********************************************************************
\author{Roberto Pittau 
% Optional short acknowledgment: remove next line if non-needed
\thanks{I thank the financial support of the MICINN project FPA2011-22398
(LHC@NLO).}
% DO NOT MODIFY THE FOLLOWING '\vspace' ARGUMENT
\vspace{.3cm}\\
% Addresses and institutions (remove "1- " in case of a single institution)
Departamento de F\'{\i}sica Te\'orica y del Cosmos \\
 Universidad de Granada - Spain
}
%%***********************************************************************
% END OF AUTHORS INFORMATION AREA
%***********************************************************************

\maketitle

\begin{abstract}
I review the present status of the automatic NLO tools
{\sc MadLoop} and a{\sc MC@NLO} by presenting, as an example of their use,
phenomenological studies of hadron collider processes.
Perspectives on applications to linear collider Physics 
are also discussed.
\end{abstract}

\section{Introduction}
In this contribution, I review the present status of the {\sc MadLoop}~\cite{Hirschi:2011pa,Hirschi:2011rb} and a{\sc MC@NLO}~\cite{amcatnlo} projects, whose aim is computing observables at the Next to Leading Order (NLO) accuracy at high energy colliders.
Both tools are based on the strategic
assumption that, for the word {\it automation} to have its
proper meaning, the only operation required from a user is that of typing-in
the process to be computed, and other analysis-related information (such as
final-state cuts). In particular, the codes that achieve the {\it automation} 
may only differentiate between processes depending on their general
characteristics, but must never work on a case-by-case basis.

In Section~\ref{Sec:1}, I review the general structure of a typical NLO calculation and the computational strategies implemented in {\sc MadLoop}.
Section~\ref{Sec:2} describes the a{\sc MC@NLO} framework while, in 
Section~\ref{Sec:3}, I report on three phenomenological studies relevant at the LHC.  
Finally, in Section~\ref{Sec:4}, I discuss the ongoing work to extend the applicability of such automatic tools to linear collider Physics.

\section{{\sc MadLoop} and the problem of computing the 1-loop corrections\label{Sec:1}}
The typical structure of a NLO calculation is given by the following formula:
\begin{eqnarray}
\label{nlocalc}
\sigma^{NLO} = \int_m d \sigma^{B} 
+\int_m \left( d \sigma^{V} + { \int_1    d \sigma^{A}}  \right)
+ \int_{m+1} \left( d \sigma^{R} - { d \sigma^{A}} \right)\,,
\end{eqnarray}
where $d \sigma^{B}$ is the Born cross section, 
$d \sigma^{V}$ is the virtual 1-loop correction, $d \sigma^{R}$ 
is the real correction, and $d \sigma^{A}$ and $\int_1 d \sigma^{A}$  are 
the {\it unintegrated} and
{\it integrated} counterterms, respectively, which allow to compute the real 
contribution in 4 dimensions.
The complexity of the calculation grows up rapidly with the number of external legs, especially in the sector of the 1-loop corrections.
Since any 1-loop amplitude $A$ can be decomposed in terms of known scalar 1-loop functions
\begin{eqnarray}
\label{eq:amp}
A  &=& 
\sum_{i_0 < i_1 < i_2 < i_3}^{m-1} {d}( i_0 i_1 i_2 i_3) 
\int d^n \bar{q} \frac{1}{\db{i_0}\db{i_1}\db{i_2}\db{i_3}} 
  + \sum_{i_0 < i_1 < i_2 }^{m-1}{ c}( i_0 i_1 i_2)
\int d^n \bar{q} \frac{1}{\db{i_0}\db{i_1}\db{i_2}} \nonumber \\
&&+\sum_{i_0 < i_1 }^{m-1} { b}(i_0 i_1) 
\int d^n \bar{q} \frac{1}{\db{i_0}\db{i_1}} 
  + \sum_{i_0}^{m-1} { a}(i_0) \int d^n \bar{q} \frac{1}{\db{i_0}} 
+R\,, 
\end{eqnarray}
the problem is fully solved once one determines the set of coefficients
\begin{eqnarray}
\label{eq:coeff}
{\cal S} = \left\{
\begin{tabular}{lll}
$\!\!{d}( i_0 i_1 i_2 i_3)$, & $\!\!{c}( i_0 i_1 i_2)$, & \\
$\!\!{b}( i_0 i_1)$,         & $\!\!{a}(i_0)$,          & 
$\!\!{R}\,.$ 
\end{tabular}
\right.
\end{eqnarray}
That is obtained in {\sc MadLoop} with the OPP approach~\cite{Ossola:2006us,Ossola:2007bb,Ossola:2008xq,Draggiotis:2009yb,Pittau:2010tk}, in which the natural object is the {\em integrand} of the virtual 1-loop amplitude
\begin{eqnarray}
A = \int 
d^d q
~{\cal A}(q)  \,.
\end{eqnarray}
For example, in the case of a $2 \to 4$ process, ${\cal A}(q)$ can be cast in the form
\begin{eqnarray}
 {\cal A}(q)= \sum \frac{N_i^{(6)}(q)}{\db{i_0}\db{i_1} \cdots \db{i_5}}
+ \sum \frac{N_i^{(5)}(q)}{\db{i_0}\db{i_1} \cdots \db{i_4}}
+ \sum \frac{N_i^{(4)}(q)}{\db{i_0}\db{i_1} \cdots \db{i_3}}
+ \cdots
\end{eqnarray}
where the $N_i^{k}(q)$ denote numerator functions of all possible structures with $k$ denominators. 
The coefficients in Equation~\ref{eq:coeff} are obtained in {\sc MadLoop} 
by sampling numerically the numerators $N_i^{k}(q)$ (considered as functions of 
the would-be loop momentum $q$) with the help of {\sc CutTools}~\cite{Ossola:2007ax}. Since $N_i^{k}(q)$  are tree-level like objects, they are suitable to be computed with {\sc MadGraph}, which is the tool used by {\sc MadLoop} for their numerical determination \footnote{A version working with {\sc MadGraph4} is available in~\cite{amcatnlo}, work is in progress for interfacing to {\sc MadGraph5}~\cite{Alwall:2011uj}.}.

\begin{figure}[ht]
\begin{center}
\includegraphics[width=0.49\textwidth]{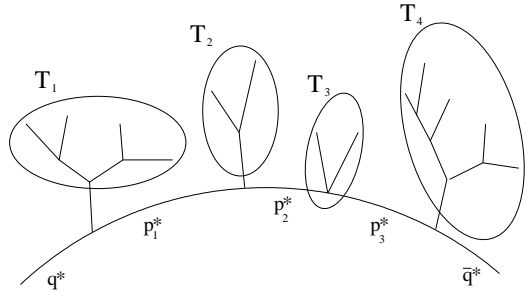}
 \caption{Example on an L-cut diagram.}
\label{openloop}
\end{center}
\end{figure}

One-loop amplitudes for $2 \to n$ processes are constructed in {\sc MadLoop}  
by sewing {\em tree-level} $2 \to n+2$ diagrams (dubbed L-cut diagrams) where 
one special line L is opened. Then amplitudes are formed by discarding one-loop diagrams in excess, a process  that we call {\em diagram filtering}. Due to the 1-loop topology, two L-cut diagrams must be considered equivalent if they are identical up to a {cyclic permutation}, or to {mirror symmetry}, or to a cyclic permutation plus mirror symmetry. An example of L-cut diagram, identified with the string 
\begin{eqnarray}
q^\star\,T_1\,p_1^\star\,T_2\,p_2^\star\,T_3\,p_3^\star\,T_4\,q^\star
\nonumber
\end{eqnarray}
is given in Figure~\ref{openloop}.

\noindent The following 2 strings are equivalent to that one depicted in Figure~\ref{openloop} 
\begin{eqnarray}
p_1^\star\,T_2\,p_2^\star\,T_3\,p_3^\star\,T_4\,q^\star\,T_1\,
p_1^\star~~~~~~~
q^\star\,T_4\,p_3^\star\,T_3\,p_2^\star\,T_2\,p_1^\star\,T_1\,q^\star\,.
\end{eqnarray}

It is easy to convince oneself that, when computing QCD corrections,
the L-cut processes one needs to consider correspond to the following
choices of the L-cut particles:
\begin{eqnarray}
(q^\star,\,\bar{q}^\star)&=&(g,\,g)
\phantom{;\,(d,\,\bar d);\ldots\,(Q,\,\bar Q)}
\;\;\;\;\;\;\;\;\,{\rm gluons}\,,
\label{qqglu}
\\
&=&(u,\,\bar u);\,(d,\,\bar d);\ldots\,(Q,\,\bar Q)
\;\;\;\;\;\;\;\;{\rm quarks}\,,
\label{qqqrk}
\\
&=&(\eta,\,\bar{\eta})
\phantom{;\,(d,\,\bar d);\ldots\,(Q,\,\bar Q)}
\;\;\;\;\;\;\;\;\,{\rm ghosts}\,.
\label{qqeta}
\end{eqnarray}
In general theories, the L-cuts processes corresponding to all particles in the Lagrangian should be taken into account.

\section{a{\sc MC@NLO}\label{Sec:2}}
a{\sc MC@NLO} is a fully automated approach to
complete event generation and subsequent parton shower at the NLO accuracy
in QCD, which allows accurate and flexible
simulations for both signals and backgrounds at hadron colliders.  All
calculational aspects in a{\sc MC@NLO} are automated.  One-loop
contributions are evaluated with {\sc MadLoop} as described in the previous section.
The other matrix-element contributions to the cross sections 
appearing in Equation~\ref{nlocalc}, their phase-space subtractions according
to the FKS formalism~\cite{Frixione:1995ms,Frixione:2011kh}, 
their combinations with the
one-loop results, and their integration are performed by
{\sc MadFKS}~\cite{Frederix:2009yq}. The matching of the NLO results with {\sc
HERWIG}~\cite{Corcella:2000bw} or {\sc PYTHIA}~\cite{Sjostrand:2006za} parton showers is performed with the MC@NLO
method~\cite{Frixione:2002ik}, and it is also completely automatic.
An important aspect of the above procedure is that all the ingredients 
of Equation~\ref{nlocalc} can be  computed  independently and put together to produce physical results in a subsequent stage \footnote{For example, conventions 
exist~\cite{Binoth:2010xt} to interface real and virtual parts of a NLO computation.} .

As an example of the use and validation of our automatic tools we report, in Table~\ref{tab:results}, parton level results obtained with {\sc MadFKS} and {\sc MadLoop}.
\begin{table}[ht]
\begin{center}
\begin{small}
\hspace{-1.cm}
%\begin{tabular}{\textwidth}
%{lr@{$\,\to\,$}lccr@{$\,\pm\,$}lr@{$\,\pm\,$}X}
\begin{tabular}{crlccrlrl}
\hline
\multicolumn{3}{c}{Process~~~~~~~~~~~~~~} & 
  $\mu$ & $n_{lf}$ &
  \multicolumn{4}{c}{Cross section (pb)}
\\
\multicolumn{3}{c}{} &&& \multicolumn{2}{c}{LO} & \multicolumn{2}{c}{NLO}
%%%%%%%%%%%%%%%%%%%%%%%%%%%%%%%%%%%%%%%%%%%%%%%%%%%%%%%%%%%%%%%%%%%%%%%%
\\
\hline
a.1 & $pp \to $ & $\!\!\!\!\!\! t\bar{t}$     & $m_{top}$   &5& $123.76\, \pm$ & $\!\!\!\!\!\!\! 0.05$ & $162.08\, \pm$ & $\!\!\!\!\!\!\! 0.12$ \\
a.2 & $pp \to $ & $\!\!\!\!\!\! tj$           & $m_{top}$   &5& $34.78\, \pm$ & $\!\!\!\!\!\!\! 0.03$ & $41.03\, \pm$ & $\!\!\!\!\!\!\! 0.07$ 
\\
a.3 & $pp \to $ & $\!\!\!\!\!\! tjj$          & $m_{top}$   &5& $11.851\, \pm$ & $\!\!\!\!\!\!\! 0.006$ & $13.71\, \pm$ & $\!\!\!\!\!\!\! 0.02$
\\
a.4 & $pp \to $ & $\!\!\!\!\!\! t\bar{b}j$    & $m_{top}/4$ &4& $25.62\, \pm$ & $\!\!\!\!\!\!\! 0.01$ & $30.96\, \pm$ & $\!\!\!\!\!\!\! 0.06$
\\
a.5 & $pp \to $ & $\!\!\!\!\!\! t\bar{b}jj$   & $m_{top}/4$ &4& $8.195\, \pm$ & $\!\!\!\!\!\!\! 0.002$ & $8.91\, \pm$ & $\!\!\!\!\!\!\! 0.01$
%%%%%%%%%%%%%%%%%%%%%%%%%%%%%%%%%%%%%%%%%%%%%%%%%%%%%%%%%%%%%%%%%%%%%%%%
\\
\hline
b.1 & $pp \to $ & $\!\!\!\!\!\! (W^+\to) e^+\nu_e$          & $m_W$ &5& $5072.5\, \pm$ & $\!\!\!\!\!\!\!2.9$ & $6146.2\, \pm$ & $\!\!\!\!\!\!\!9.8$ 
\\
b.2 & $pp \to $ & $\!\!\!\!\!\! (W^+\to) e^+\nu_e\,j$       & $m_W$ &5& $828.4\, \pm$ & $\!\!\!\!\!\!\!0.8$ & $1065.3\, \pm$ & $\!\!\!\!\!\!\!1.8$  
\\
b.3 & $pp \to $ & $\!\!\!\!\!\! (W^+\to) e^+\nu_e\,jj$      & $m_W$ &5& $298.8\, \pm$ & $\!\!\!\!\!\!\!0.4$ & $300.3\, \pm$ & $\!\!\!\!\!\!\!0.6$  
\\
b.4 & $pp \to $ & $\!\!\!\!\!\! (\gamma^*/Z\to) e^+e^-$     & $m_Z$ &5& $1007.0\, \pm$ & $\!\!\!\!\!\!\!0.1$ & $1170.0\, \pm$ & $\!\!\!\!\!\!\!2.4$ 
\\
b.5 & $pp \to $ & $\!\!\!\!\!\! (\gamma^*/Z\to) e^+e^-\,j$  & $m_Z$ &5& $156.11\, \pm$ & $\!\!\!\!\!\!\!0.03$ & $203.0\, \pm$ & $\!\!\!\!\!\!\!0.2$ 
\\
b.6 & $pp \to $ & $\!\!\!\!\!\! (\gamma^*/Z\to) e^+e^-\,jj$ & $m_Z$ &5& $54.24\, \pm$ & $\!\!\!\!\!\!\!0.02$ & $56.69\, \pm$ & $\!\!\!\!\!\!\!0.07$ 
%%%%%%%%%%%%%%%%%%%%%%%%%%%%%%%%%%%%%%%%%%%%%%%%%%%%%%%%%%%%%%%%%%%%%%%%
\\
\hline
c.1 & $pp \to $ & $\!\!\!\!\!\! (W^+\to) e^+\nu_e b\bar{b}$      & $m_W+2m_b$ &4& $11.557\, \pm$ & $\!\!\!\!\!\!\!0.005$ & $22.95\, \pm$ & $\!\!\!\!\!\!\!0.07$ 
\\
c.2 & $pp \to $ & $\!\!\!\!\!\! (W^+\to) e^+\nu_e t\bar{t}$      & $m_W+2m_{top}$ &5& $0.009415\, \pm$ & $\!\!\!\!\!\!\!0.000003$ & $0.01159\, \pm$ & $\!\!\!\!\!\!\!0.00001$ 
\\
c.3 & $pp \to $ & $\!\!\!\!\!\! (\gamma^*/Z\to) e^+e^- b\bar{b}$ & $m_Z+2m_b$ &4& $9.459\, \pm$ & $\!\!\!\!\!\!\!0.004$ & $15.31\, \pm$ & $\!\!\!\!\!\!\!0.03$ 
\\
c.4 & $pp \to $ & $\!\!\!\!\!\! (\gamma^*/Z\to) e^+e^- t\bar{t}$ & $m_Z+2m_{top}$ &5& $0.0035131\, \pm$ & $\!\!\!\!\!\!\!0.0000004$ & $0.004876\, \pm$ & $\!\!\!\!\!\!\!0.000002$ 
\\
c.5 & $pp \to $ & $\!\!\!\!\!\! \gamma t\bar{t}$                & $2m_{top}$     &5& $0.2906\, \pm$ & $\!\!\!\!\!\!\!0.0001$ & $0.4169\, \pm$ & $\!\!\!\!\!\!\!0.0003$ 
%%%%%%%%%%%%%%%%%%%%%%%%%%%%%%%%%%%%%%%%%%%%%%%%%%%%%%%%%%%%%%%%%%%%%%%%
\\
\hline
d.1 & $pp \to $ & $\!\!\!\!\!\! W^+W^-$     & $2m_W$ &4& $29.976\, \pm$ & $\!\!\!\!\!\!\!0.004$ & $43.92\, \pm$ & $\!\!\!\!\!\!\!0.03$ 
\\
d.2 & $pp \to $ & $\!\!\!\!\!\! W^+W^-\,j$  & $2m_W$ &4& $11.613\, \pm$ & $\!\!\!\!\!\!\!0.002$ & $15.174\, \pm$ & $\!\!\!\!\!\!\!0.008$ 
\\
d.3 & $pp \to $ & $\!\!\!\!\!\! W^+W^+\,jj$ & $2m_W$ &4& $0.07048\, \pm$ & $\!\!\!\!\!\!\!0.00004$ & $0.1377\, \pm$ & $\!\!\!\!\!\!\!0.0005$ 
%%%%%%%%%%%%%%%%%%%%%%%%%%%%%%%%%%%%%%%%%%%%%%%%%%%%%%%%%%%%%%%%%%%%%%%%
\\
\hline
e.1 & $pp \to $ & $\!\!\!\!\!\! HW^+$        & $m_W+m_H$ &5& $0.3428\, \pm$ & $\!\!\!\!\!\!\!0.0003$ & $0.4455\, \pm$ & $\!\!\!\!\!\!\!0.0003$ 
\\
e.2 & $pp \to $ & $\!\!\!\!\!\! HW^+\,j$     & $m_W+m_H$ &5& $0.1223\, \pm$ & $\!\!\!\!\!\!\!0.0001$ & $0.1501\, \pm$ & $\!\!\!\!\!\!\!0.0002$ 
\\
e.3 & $pp \to $ & $\!\!\!\!\!\! HZ$        & $m_Z+m_H$ &5& $0.2781\, \pm$ & $\!\!\!\!\!\!\!0.0001$ & $0.3659\, \pm$ & $\!\!\!\!\!\!\!0.0002$ 
\\
e.4 & $pp \to $ & $\!\!\!\!\!\! HZ\,j$     & $m_Z+m_H$ &5& $0.0988\, \pm$ & $\!\!\!\!\!\!\!0.0001$ & $0.1237\, \pm$ & $\!\!\!\!\!\!\!0.0001$ 
\\
e.5 & $pp \to $ & $\!\!\!\!\!\! Ht\bar{t}$ & $m_{top}+m_H$ &5& $0.08896\, \pm$ & $\!\!\!\!\!\!\!0.00001$ & $0.09869\, \pm$ & $\!\!\!\!\!\!\!0.00003$ 
\\
e.6 & $pp \to $ & $\!\!\!\!\!\! Hb\bar{b}$ & $m_b+m_H$ &4& $0.16510\, \pm$ & $\!\!\!\!\!\!\!0.00009$ & $0.2099\, \pm$ & $\!\!\!\!\!\!\!0.0006$ 
\\
e.7 & $pp \to $ & $\!\!\!\!\!\! Hjj$       & $m_H$     &5& $1.104\, \pm$ & $\!\!\!\!\!\!\!0.002$ & $1.036\, \pm$ & $\!\!\!\!\!\!\!0.002$
\\
\hline
\end{tabular}
\end{small}
\end{center}
\caption{\label{tab:results}
Results for total rates, possibly within cuts, at the 7~TeV LHC,
obtained with {\sc MadFKS} and {\sc MadLoop}. The errors are due to the statistical
uncertainty of Monte Carlo integration.}
\end{table}

\section{Results at the LHC\label{Sec:3}}
To illustrate the kind of realistic analyses one can perform, I list here 
the a{\sc MC@NLO} predictions for the processes $pp \to ttH$~\cite{Frederix:2011zi}, $pp \to Vbb$~\cite{Frederix:2011qg} and $pp \to \ell^+\ell^- \ell^{(\prime)+} \ell^{(\prime)-}$~\cite{Frederix:2011ss} at the 7~TeV LHC. Notice that an automatic procedure to determine scale and PDF uncertainties is available in the a{\sc MC@NLO} framework, which allows to estimate them at almost zero CPU cost by a simple reweighting procedure. Results obtained with such a method are presented 
in Table~\ref{amcatnlotab1} of Subsection~\ref{4leptons}.
\subsection{The $ttH$ process}
\begin{figure}[ht]
\begin{center}
\includegraphics[width=0.49\textwidth]{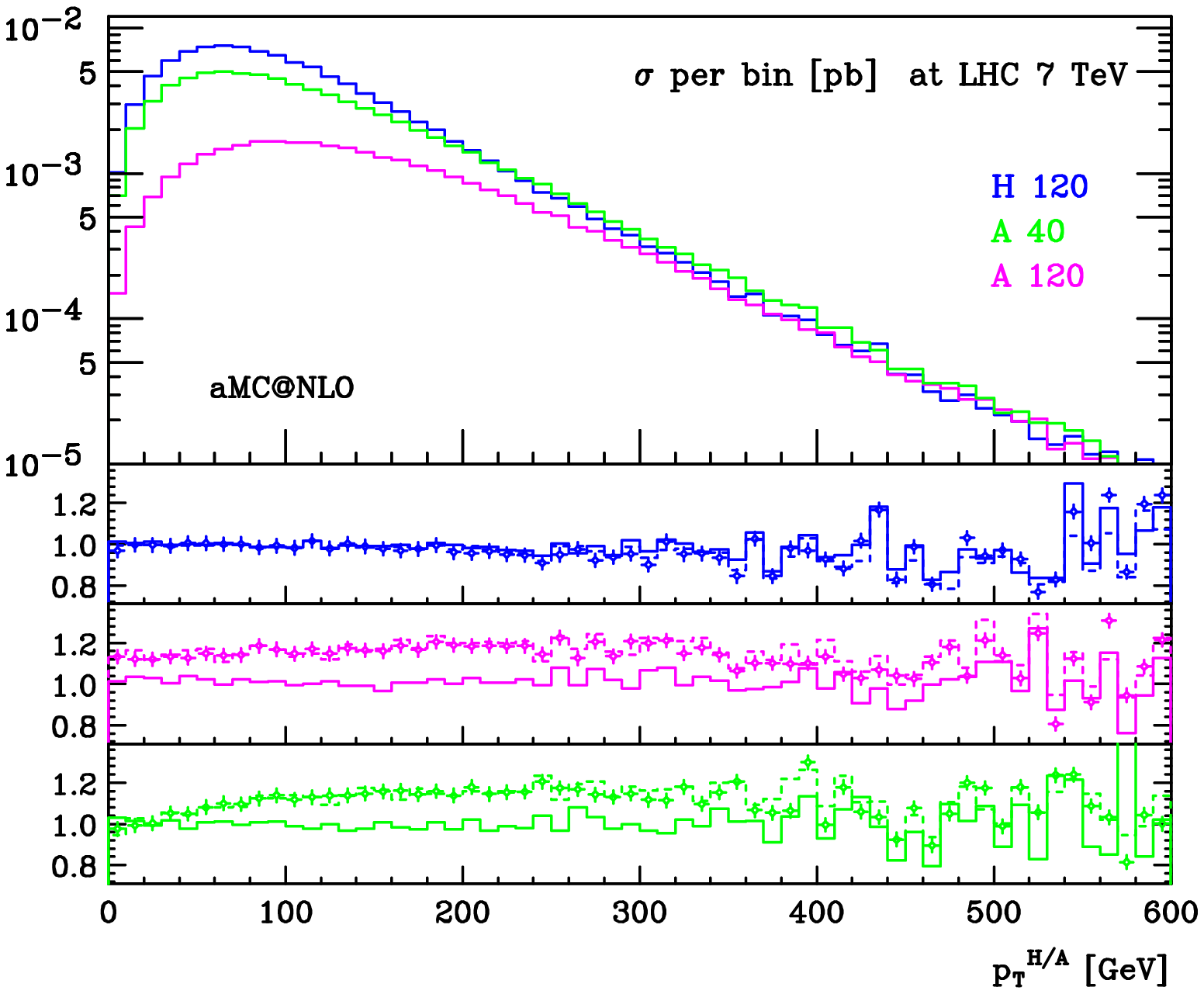}
\includegraphics[width=0.49\textwidth]{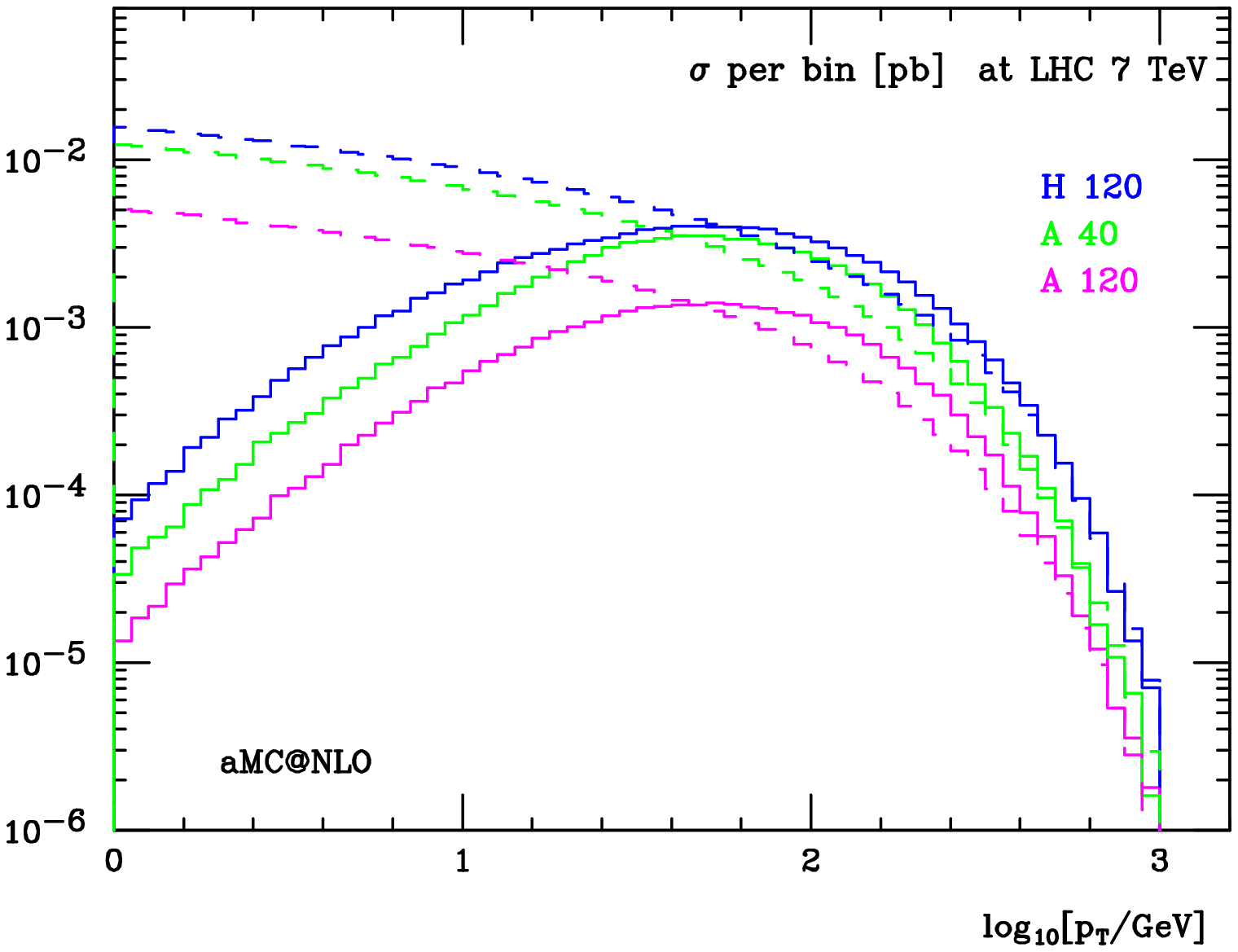}
 \caption{Higgs transverse momentum distributions (left) and 
    transverse momentum of the $ttH$ or $ttA$ system (right)
    in $ttH/ttA$ events at the LHC ($\sqrt{s}$=7 TeV), 
    with a{\sc MC@NLO} in the three cases: Scalar (blue) and pseudoscalar 
    (magenta)
    Higgs with $m_{H/A}=120$ GeV and pseudoscalar (green) with $m_A=40$ GeV. In
    the lower panels of the left part, the ratios of a{\sc MC@NLO} over LO 
    (dashed), NLO (solid),
    and aMC@LO (crosses) are shown. Solid histograms in the right panel
    are relevant to a{\sc MC@NLO}, dashed ones to a pure NLO calculation.}
\label{amcatnlofig1}
\end{center}
\end{figure}

The production process of a $H$ boson in association 
with a top pair is a classic mechanism for Higgs production at the 
LHC~\cite{Dittmaier:2011ti,LHCHiggsCrossSectionWorkingGroup:2012vm}, 
where the large $ttH$ Yukawa coupling and the presence of top quarks can be exploited to extract the signal from its QCD multi-jet background.
As an example of the use of  a{\sc MC@NLO} for this process we present, in
Figure~\ref{amcatnlofig1}, the Higgs transverse momentum distribution 
and the transverse momentum of the $ttH$ or $ttA$ system 
for a standard model (scalar) Higgs with 
$M_H=$ 120 GeV and for a pseudoscalar one with $M_A=$ 120/40 GeV.
The total NLO cross sections in the three cases are 
$\sigma_{\rm NLO}(M_H= 120)=$ 103.4 fb,
$\sigma_{\rm NLO}(M_A= 120)=$ 31.9 fb,
and 
$\sigma_{\rm NLO}(M_A=  40)=$ 77.3 fb, respectively.
At moderate values of the Higgs transverse momentum, the scalar and
pseudoscalar cases are clearly distinguishable, while at larger values the
three distributions tend to coincide. Parton shower effects give in general
small corrections with respect to the a pure NLO calculation, except for
variables involving all produced particles, such as the transverse momentum of
the $ttH$ or $ttA$ system shown in the right panel of
Figure~\ref{amcatnlofig1}.

\subsection{The $Vbb$ process}
\begin{figure}[ht]
\begin{center}
\includegraphics[width=0.5\textwidth]{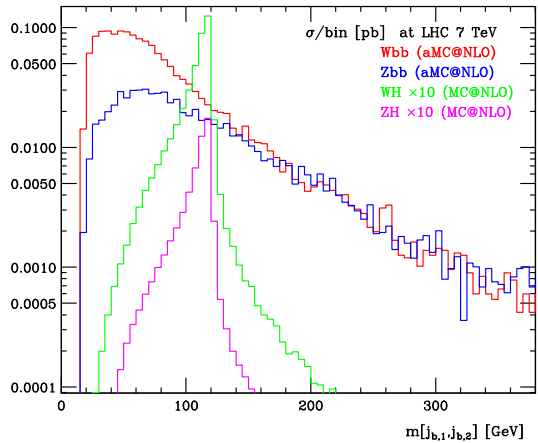}
 \caption{Invariant mass of the pair of the two leading $b$-jets.
$WH(\to \ell\nu bb)$, $ZH(\to\ell^+\ell^- bb)$, $\ell \nu bb$,
and $\ell^+ \ell^- bb$ results are shown, with the former two rescaled by a factor of ten.}
\label{amcatnlofig2a}
\end{center}
\end{figure}

With $Vbb$ we understand $\ell \nu bb$ and
$\ell^+ \ell^- bb$ final states, which are the main backgrounds to searches for 
SM Higgs production in association with
vector bosons ($WH/ZH$), with the subsequent Higgs decay into 
a $bb$ pair.
The a{\sc MC@NLO} framework allows a realistic study including
\begin{itemize}
\item NLO corrections;
\item bottom quark mass effects;
\item spin-correlation and off-shell effects;
\item showering and hadronization.
\end{itemize}
As an example we show, in Figure~\ref{amcatnlofig2a}, the invariant mass of the pair of the two leading b-jets, compared with the signal distributions for a standard Higgs with $m_H = 120$ GeV. Figure~\ref{amcatnlofig2a} is interesting 
because both signal and background are studied at the NLO accuracy.

\subsection{Four-lepton production\label{4leptons}}
\begin{figure}[ht]
\begin{center}
\includegraphics[width=0.49\textwidth]{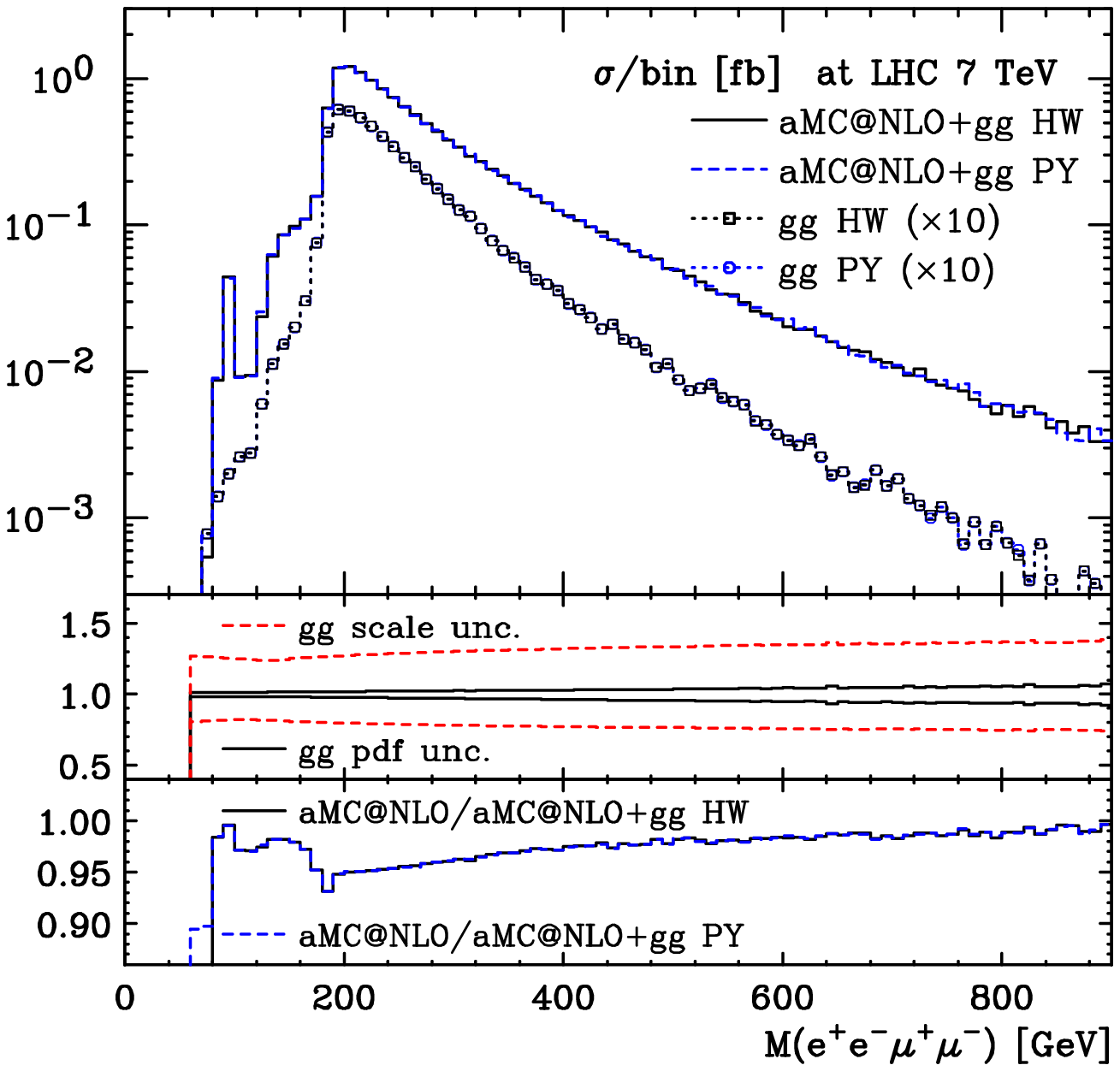}
\includegraphics[width=0.49\textwidth]{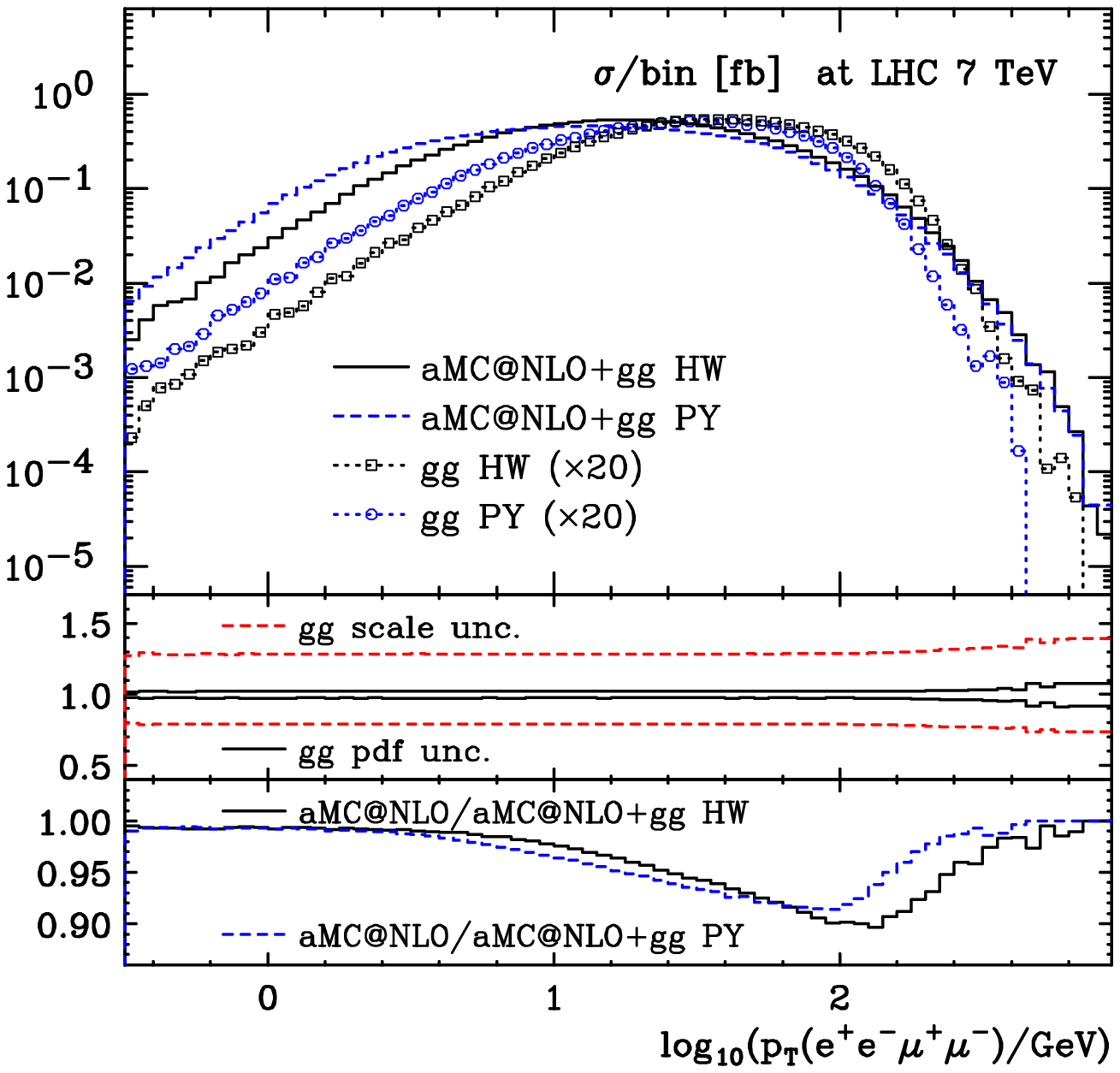}
 \caption{Four-lepton invariant mass and the transverse momentum distributions 
for a{\sc MC@NLO} +$gg$
{\sc HERWIG} (solid black) and {\sc PYTHIA} (dashed blue) results. The rescaled $gg$
contributions with {\sc HERWIG} (open black boxes) and {\sc PYTHIA} (open blue circles)
are shown separately. Middle insets: scale (dashed red) and PDF (solid black)
fractional uncertainties. Lower insets: a{\sc MC@NLO}/(a{\sc MC@NLO}+$gg$) with 
{\sc HERWIG} (solid black) and {\sc PYTHIA} (dashed blue).}
\label{amcatnlofig3}
\end{center}
\end{figure}

Vector boson pair production is important in at least two respects. Firstly,
it is an irreducible background to Higgs signals, in particular through the
$W^+W^-$ and ZZ channels which are relevant to searches for a
standard model Higgs of mass larger than about 140 GeV. Secondly, di-boson
cross sections are quite sensitive to violations of the gauge structure of the
standard model, and hence are good probes of scenarios where new Physics is
heavy and not directly accessible at the LHC, yet the couplings in the vector
boson sector are affected.
The neutral process
$$
pp \to (Z/\gamma^\ast) (Z/\gamma^\ast) \to  
\ell^+\ell^- \ell^{(\prime)+} \ell^{(\prime)-}\,
$$
is considered here, which, although smaller than the $W^+W^-$ channel, may provide a cleaner signal due to the possibility of fully reconstructing the decay
products of the two vector bosons. a{\sc MC@NLO} predictions for the cross
sections are given in Tab.~\ref{amcatnlotab1}, which, as already mentioned, also includes a{\sc MC@NLO} estimates for scale and PDF uncertainties.  The four-lepton invariant mass and the transverse momentum distribution are presented in
Figure~\ref{amcatnlofig3}, where comparisons between the results obtained with
a{\sc MC@NLO} matched to {\sc HERWIG} and to {\sc PYTHIA} are also given.
I stress that these results include the contributions due to 
$gg$-initiated processes, which have also been computed automatically.
These are formally of NNLO, but may play a non-negligible phenomenological
role owing to their parton-luminosity dominance at a large-energy
collider such as the LHC.
%%%%%%%%%%%%%%%%%%%%%%%%%%%%%%%%%%%%%%%%%%%%%%%%%%%%%%%%%%%%%%%%%%%%%%%%
\begin{table}[ht]
\begin{center}
\renewcommand*{\arraystretch}{1.5}
\begin{tabular}{|c|cc|c|}
\hline
 & \multicolumn{3}{c|}{Cross section (fb)}\\
\cline{2-4}  
 Process  & \multicolumn{2}{c|}{$q\bar q$/$qg$ channels} & {$gg$ channel} \\
\cline{2-4}
& ${\cal O}(\alpha_{\scriptscriptstyle s}^0)$ & ${\cal O}(\alpha_{\scriptscriptstyle s}^0)+{\cal O}(\alpha_{\scriptscriptstyle s})$ & ${\cal O}(\alpha_{\scriptscriptstyle s}^2)$ \\
\hline
$pp\to e^+e^-\mu^+\mu^-$&   
% 9.193 & $12.90^{+0.27(2.1\%)+0.26(2.0\%)}_{-0.23(1.8\%)-0.22(1.7\%)}$  & 
 9.19 & $12.90^{+0.27(2.1\%)+0.26(2.0\%)}_{-0.23(1.8\%)-0.22(1.7\%)}$  & 
 $0.566^{+0.162(28.6\%)+0.012(2.1\%)}_{-0.118(20.8\%)-0.014(2.5\%)}$     \\
$pp \to e^+ e^- e^+ e^-$      &   
%% 4.577 & $6.435^{+0.133(2.1\%)+0.111(1.7\%)}_{-0.130(2.0\%)-0.100(1.6\%)}$&\\
 4.58 & $6.43^{+0.13(2.1\%)+0.11(1.7\%)}_{-0.13(2.0\%)-0.10(1.6\%)}$ & \\
\hline
\end{tabular}
\end{center}
\caption{Total cross sections for $e^+e^-\mu^+\mu^-$ and $e^+e^-e^+e^-$
production at the LHC ($\sqrt{S}= 7$~TeV) within the cuts 
$M(\ell^\pm\ell^{(\prime)\mp})\ge 30~{\rm GeV}$. The first and second errors 
affecting the results
are the scale and PDF uncertainties (also given as fractions of the
central values).}
\label{amcatnlotab1}
\end{table}
%%%%%%%%%%%%%%%%%%%%%%%%%%%%%%%%%%%%%%%%%%%%%%%%%%%%%%%%%%%%%%%%%%%%%%%%

\section{Perspectives for linear collider Physics\label{Sec:4}}
Automatic tools such as those described in the previous sections can be successfully employed for studying ILC/CLIC $e^+e^-$ Physics as well.
 However, for this to be achieved in practice, a number of intermediate technical challenges should be dealt with. If, on the one hand, the treatment of the real radiation is much easier due to the much simpler initial state, on the other hand, the need of including the full set of electroweak 1-loop corrections could pose a problem in terms of speed due to the much larger number of contributing Feynman diagrams. If, in addition, one wishes to study SUSY or generic BSM models at linear colliders including radiative corrections, things can become too slow.
 However, one should not forget that the intrinsic simplicity of the algorithms used in {\sc MadLoop} and a{\sc MC@NLO} is, at a very large extent, independent on the complexity of the process under study and that the results are always guaranteed to be correct thanks to the complete automation of the whole procedure. Therefore, simple cashing strategies and a clever organization of the calculation can be used to deal with such CPU challenging computations.
 Finally, the fact that both {\sc MadLoop} and a{\sc MC@NLO} are embedded into the {\sc MadGraph} framework guarantees a dedicated and constant work in the direction of extending them to theories more complicated than QCD. For example, the 
complete electroweak standard model in the renormalizable gauge is being currently implemented in {\sc MadGraph5}, together with the ultraviolet counterterms and the needed so called $R_2$ finite renormalization~\cite{Garzelli:2009is,Garzelli:2010qm}. Work is also in progress for SUSY and BSM theories~\cite{Christensen:2008py,Degrande:2011ua,Pittau:2011qp}.
\section{Conclusion}
Physics at the Next-to-Leading order accuracy should be easy and the tools dedicated to its study at high-energy colliders user friendly. 
In order to allow a quick progress in our understanding of the fundamental laws of nature, the human effort should be better employed to compare easily produced theoretical predictions with data. 
\begin{footnotesize}
% IF YOU DO NOT USE BIBTEX, USE THE FOLLOWING SAMPLE SCHEME FOR THE REFERENCES
% ----------------------------------------------------------------------------

\end{footnotesize}

\end{document}